\documentclass{article}
\title{\bf One-loop quantum cosmological correction to the gravitational constant using the kink solution in de Sitter universe}
\author{F. Darabi\thanks{e-mail:
C.Author-f.darabi@azaruniv.edu} and S. Jalalzadeh$^1$\thanks{email: s-jalalzadeh@sbu.ac.ir} \\{\small Department of Physics, Azarbaijan University of
Tarbiat Moallem, 53714-161, Tabriz, Iran .}\\{\small Department of
Physics, Shahid Beheshti University, Evin, Tehran 19839,
Iran} }
\begin{document}
\maketitle
\begin{abstract}
In this paper, we show the equivalence between a classical static scalar field theory and the (closed) de Sitter cosmological model whose potential represents shape invariance property. Based on this equivalence, we calculate the one-loop quantum cosmological correction to the ground state energy of the kink-like solution in the (closed) de Sitter cosmological model in which the fluctuation potential $V^{\prime\prime}$ has a shape invariance property. It is shown that this type of correction, which yields a renormalized mass in the case of scalar field theory, may
be {\it interpreted} as a renormalized gravitational constant in the case of (closed) de Sitter cosmological model.
\end{abstract}
\newpage
\section{Introduction}

The quantum corrections to the mass of classical topological defects
plays an important role in the semi-classical approach to quantum
field theory \cite{1,1'}. The computation of quantum energies around
classical configurations in (1 + 1)-dimensional kinks has been
developed in \cite{2,2',2'',2'''} by using topological boundary conditions, the
derivative expansion method \cite{3,3',3'',3'''',3'''}, the scattering phase shift
technique \cite{2'}, the mode regularization approach \cite{5}, the
zeta- function regularization technique \cite{14.14,6}, and also the
dimensional regularization method \cite{7}. In a previous paper by
one of the authors, the one-loop renormalized kink quantum mass
correction in a (1 + 1)-dimensional scalar field theory model was
derived using the generalized zeta function method for those
potentials where the fluctuation potential $V^{''}$ has the shape
invariance property \cite{8}. These potentials are very important
since they possess a shape invariant operator in their prefactor
which makes the corrections of this kind of potential exact by the
heat kernel method. This kind of potential occurs in different
fields of physics particularly in quantum gravity and cosmology. An
empty closed Friedmann-Robertson-Walker (FRW) universe with a
decaying cosmological term $\Lambda \sim a^{-m}$ is an example ($a$
is the scale factor and $m$ is a parameter $0\leq m \leq 2$) which
is equivalent to a cosmology with the exotic matter equation of
state $p=(m/3-1)\rho$ \cite{9}. The special case $m=0$ leads to de
Sitter spacetime which is of particular interest in the present
work.

One of the basic motivations to study the quantum gravity on de Sitter background
was the fact that it might be a very reasonable candidate for vacuum state when the classical action contains a positive cosmological term \cite{1.1,2.2,3.3,4.4,5.5,6.6,7.7,8.8,9.9,9.9',10.10,11.11,12.12}. Moreover, such investigations could provide the framework to solve the
cosmological constant problem \cite{1.1,2.2,4.4,5.5,11.11}. A quite interesting
model in this regard was the Coleman-Weinberg type suppression for the effective cosmological constant using the large-distance limit of the quantum gravity one-loop effective action on the de
Sitter background \cite{12.12}. In order to slightly improve the situation
such background was replaced by an hyperbolic background (i.e. a gravitational theory with negative cosmological constant) \cite{13.13}. The one-loop effective action for 4-dimensional gauged supergravity with negative cosmological
constant, was also investigated in space-times with compact hyperbolic spatial section. The explicit expansion of the effective action as a power series of the curvature on hyperbolic background was derived, making use of heat-kernel and zeta-regularization techniques \cite{14.14}, and the induced cosmological and Newton constants were computed \cite{15.15}. It is also worth noticing
to numbers of interesting works where quantum vacuum energy in quantum gravity on de Sitter background has been calculated \cite{16.16,17.17,18.18}.

On the other hand, we know that quantum cosmology is a rather simplified
and approximate version of quantum gravity where the key role is played by
the quantum Wheeler-DeWitt equation. In fact, quantum cosmology is just the quantum mechanics of the whole universe, not a quantum filed theory of gravity.
Therefore, for the same reason that we are interested in calculation of the vacuum energy and its corrections in quantum gravity on de Sitter background, we may study the vacuum energy and its corrections in quantum cosmology in de Sitter universe. However, we have
to keep in mind that these approximate corrections are just valuable in the context of quantum cosmology, not quantum gravity, and any nontrivial result
obtained in this way should be interpreted quantum mechanically and not quantum
field theoretically. For example, one may study the impact of these quantum
cosmological corrections on the quantum tunneling rate from {\it nothing}
to de Sitter universe \cite{Darabi}. The type of these corrections, however, depends on the
way we implement our quantum mechanical approximation. 

In this paper, we aim to make maximum use of the shape invariance property
of the quantum mechanical potential which appears in the quantum cosmology
(Wheeler-DeWitt equation) 
of de Sitter universe. In this regard, we first show the equivalence between the configuration space of (closed) de Sitter cosmological model and a {\it
classical} static scalar field theory. As mentioned above, we know how to compute the quantum corrections to the vacuum energy
of the classical scalar field, which is interpreted as the quantum corrections to the
mass of the classical kink solution. Based on this equivalence, we implement the above mentioned techniques to calculate the one-loop quantum cosmological correction to the ground state energy of the kink-like solution in the (closed) de Sitter cosmological model\footnote{It is worth noticing that usually it is not necessary in quantum gravity to use the kink solution for calculation of vacuum energy, but here in quantum cosmology we are making use of the shape invariance property and the equivalence between the configuration space of (closed) de Sitter cosmological model and a {\it
classical} static scalar field theory, so we shall use the technique where kink solution is involved to compute the quantum energies around the classical kink-like configurations in de Sitter universe.}. 
To this end, we first study the
quantum cosmology of de Sitter spacetime and construct the Euclidean
action in the path integral. Then, by a technic known as
Duru-Kleinert equivalency we change this action into the standard
quadratic one with shape invariant potential. Using the shape
invariance property of the potential, the heat kernel method, and the generalized zeta function regularization method to implement our setup for describing semi-classical kink-like states, we obtain the one-loop quantum correction to the corresponding mass-like term in the kink-like solution of the cosmological model at hand. It is shown that this type of one-loop correction, which yields quantum correction to the mass of the kink solution in the case of classical scalar field theory, may be {\it interpreted} as a renormalized gravitational constant in the case of (closed) de Sitter cosmological model.

\section{Quantum cosmology of de Sitter universe}

We shall consider an empty closed ($k=1$)
FRW universe with a non vanishing cosmological constant
$\Lambda$. The line element is given by
\begin{eqnarray}\label{b}
ds^2=-N(t)^2c^2dt^2+a(t)^2\left[\frac{dr^2}{1-r^2}+r^2d\Omega^2\right],
\end{eqnarray}
where $a(t)$ as the scale factor is the only dynamical degree of
freedom and the laps function $N(t)$ is a pure gauge variable. The pure gravitational
action corresponding to (\ref{b}) is
\begin{eqnarray}\label{2}
\begin{array}{cc}
S=\frac{c^4}{16\pi G}\int_{\mathcal
M}d^4x\sqrt{-g}(R-2\Lambda)+\frac{c^2}{8\pi G}\int_{\partial{\mathcal
M}}d^3x\sqrt{g^{(3)}}K\\
\\
=\frac{3c^2\pi}{4G}\int\left(-\frac{a\dot{a}^2}{N}+c^2Na-\frac{\Lambda}{3}Na^3\right)dt:=
\int dt L.
\end{array}
\end{eqnarray}
In equation (\ref{2}) $\Lambda$ is the cosmological constant, ${\cal
M}={\cal R}\times S^3$ is the spacetime manifold, $K$ is the trace
of the extrinsic curvature of the spacelike boundary $\partial {\cal
M}=S^3$ and the lagrangian is defined as
\begin{eqnarray}\label{lag}
L=\frac{3c^2\pi}{4G}\left(-\frac{a\dot{a}^2}{N}+c^2Na-\frac{\Lambda}{3}Na^3\right).
\end{eqnarray}
The Hamiltonian corresponding to (\ref{lag}) is obtained
\begin{eqnarray}\label{3}
H=\Pi_a \dot{a}-L=-N\frac{G}{3\pi c^2a}\Pi^2_a-\frac{3c^2\pi}{4G}(c^2Na-\frac{\Lambda}{3}Na^3),
\end{eqnarray}
where $\Pi_a=-2a\dot{a}/N$ is the canonical momentum conjugate to
$a$. Also it is assumed in
(\ref{3}) that $\Lambda>0$. Gauge invariance of action (\ref{2})
yields the Hamiltonian constraint
\begin{eqnarray}\label{4}
-\frac{\partial H}{\partial
N}=\frac{1}{2ma}\Pi^2_a+\frac{1}{2}mc^2a-\frac{m\Lambda}{6}a^3=0,
\end{eqnarray}
where $m=3\pi c^2/(2G)$ is interpreted as the {\it mass equivalent} of de Sitter universe. The Hamiltonian constraint requires a gauge fixing condition.
Choosing ${N}=1$ , the time variable $t$ becomes essentially the
proper time. In this gauge the solution of the classical equations
of motion is given by
\begin{eqnarray}\label{5}
a(t)=a_0c\cosh(\frac{t}{a_0}),
\end{eqnarray}
where $a_0^2=3/\Lambda$ is interpreted as the minimum radius of the
universe after tunneling from nothing and the initial conditions
$a(0)=a_0$ and $\dot{a}(0)=0$ are used. This solution corresponds to
a usual de Sitter spacetime where a phase of contraction from
infinitely past time is followed by an expansion phase where the
scale factor has reached its minimum $a_0$. However, a different
cosmological scenario can give rise to an identical de Sitter
expansion if we consider analytic continuation $\tau=it+(\pi/2)a_0$.
Then one can obtain the instanton solution as
\begin{eqnarray}\label{6}
a_E=ca_0\sin(\frac{\tau}{a_0}),
\end{eqnarray}
so that the instanton is a $4$-sphere of radios $a_0$ if
$\tau/a_0\in \{-\pi/2,\pi/2\}$, with spherical three-dimensional
sections labelled by the latitude angle $\theta=a_0 \tau$. Both
metrics are related by the analytic continuation into the complex
plane of the Euclidean ``time'' $\tau$
\begin{eqnarray}\label{7}
\tau=\frac{\pi}{2}a_{0}+it,\hspace{.5cm}
a=a_{E}(\frac{\pi}{2}a_0+it),
\end{eqnarray}
which is a Wick rotation with respect to the point $\tau=(\pi/2)a_0$
in this plane. This analytic continuation can be interpreted as a
quantum nucleation of the Lorentzian de Sitter spacetime from the
Euclidean hemisphere as a matching of the two manifolds across the
equatorial section $\tau=(\pi/2)a_0,\, (t=0)$, the bounce surface of
zero extrinsic curvature. Canonical quantization of this simple
cosmological model in the coordinate representation is accomplished
by the operator realizations
\begin{eqnarray}\label{8}
a=a, \hspace{.5cm} \Pi_a=-i\hbar\frac{\partial}{\partial a}.
\end{eqnarray}

The Hamiltonian constraint becomes the Wheeler-DeWitt equation for
the wave function of the de Sitter spacetime
\begin{eqnarray}\label{9}
\left(-\frac{\hbar^2}{2m}\frac{\partial^2}{\partial
a^2}+\frac{1}{2}mc^2a^2-\frac{m}{2}\frac{a^4}{a_0^2}\right)\Psi(a)=0,
\end{eqnarray}
where the operator ordering in the kinetic term is neglected. The
corresponding one-dimensional potential
\begin{eqnarray}\label{10}
V(a)=\frac{1}{2}mc^2a^2-\frac{m}{2}\frac{a^4}{a_0^2},
\end{eqnarray}
is unbounded from below, and consequently the time of flight of a
classical particle from the largest turning point to infinity is
finite, namely
\begin{eqnarray}\label{11}
\int_{x_0>a_0}^{\infty}\frac{dx}{\sqrt{|V(x)|}}<\infty.
\end{eqnarray}
This may be in contraction with (\ref{5}) at first sight, because
this equation implies $t\rightarrow\infty$ as $a\rightarrow\infty$.
However, this is only an apparent contradiction. If we use
``conformal time'' gauge $N=a(t)$, then the kinetic terms have the
standard form.
The  relation between the conformal time and proper time
is
\begin{eqnarray}\label{13}
d\eta=\frac{1}{\sqrt{m}a(t)}dt,
\end{eqnarray}
and consequently the classical solution in the conformal gauge is
\begin{eqnarray}\label{14}
a(\eta)=\frac{a_0c}{\cos(c\sqrt{m}\eta)}.
\end{eqnarray}
It is then clear that the ``particle'' reaches infinity indeed after
finite conformal time, $\eta=\pi/(2c\sqrt{m})$. The corresponding Euclidean solution
will be
\begin{eqnarray}\label{ec}
a(\tau)=\frac{a_0c}{cosh(c\sqrt{m}\tau)}.
\end{eqnarray}

The action (\ref{2}) is
invariant under the transformations
\begin{eqnarray}\label{15}
\begin{array}{cc}
\delta a=\epsilon(t)\{a,H\}, \\
\\
\delta \Pi_a=\epsilon(t)\{\Pi_a,H\},\\
\\
\delta N=\dot\epsilon(t),
\end{array}
\end{eqnarray}
provided the parameters  vanish at the end points. This
transformations defines a ``gauge'' equivalence class of histories.
The path integral in the Euclidian region for the propagator
amplitude between fixed initial and final configurations can be
written as
\begin{eqnarray}\label{16}
\begin{array}{cc}
(a_f|a_i)=\int_0^{\infty}dN<a_f,N|a_i,0>= \int_0^\infty dN\int
D\mu[a]e^{-S_E(a,N)},
\end{array}
\end{eqnarray}
where $<a_f,N|a_i,0>$ is a Green function for the WDW equation and
the Euclidian action $S_E$ is defined in the gauge $\dot{N}=0$ as
\begin{eqnarray}\label{17}
S_E=\frac{m}{2}\int_0^1 d\tau
Na^{-1}\left(\frac{\dot{a}^2}{N^2a^{-2}}+a^2(c^2-\frac{a^2}{a_0^2})\right),
\end{eqnarray}
and integration measure in one-loop approximation is given by
\begin{eqnarray}\label{18}
D\mu[a]=\prod_t da(t)(\frac{2a(t)}{N(t)})+O(\hbar).
\end{eqnarray}
The Euclidean action (\ref{17}) is not suitable to be used in
instanton calculation techniques. The reason is that the kinetic
term is not in its standard quadratic form. It has been shown that
in such quantum cosmological model one may use the Duru-Kleinert equivalence
to work with the standard form of the action \cite{13}, \cite{12}.
Using this procedure, we find the Duru-Kleinert equivalent path
integral in the present quantum cosmological model as follows
\begin{eqnarray}\label{19}
(a_f|a_i)=\int_0^{\infty}dN\int D[a]e^{-S_0},
\end{eqnarray}
where $S_0$ has the standard quadratic form
\begin{eqnarray}\label{20}
S_0=\frac{m}{2}\int_{\tau_0}^{\tau_f}d\tau\left(\dot{a}^2+a^2c^2-\frac{a^4}{a_0^2}\right).
\end{eqnarray}

\section{Semi-Classical soliton States}
In this section, we quote briefly how one can calculate the zeta
function of an operator through the heat kernel method, in the case of scalar
field $\Phi(x)$.
Classical configuration space is found by static configuration
$\Phi(x)$, so that the energy functional
\begin{eqnarray}\label{21}
E[\Phi]=\int dx\left[\frac{1}{2}\Phi_{,\mu}
\Phi^{,\mu}+V(\Phi)\right],
\end{eqnarray}
is finite. One can describe quantum evolution in  Schrodinger
picture by the following functional equation
\begin{eqnarray}\label{22}
i\hbar\frac{\partial}{\partial t}\Phi[\phi(x),t]=H\Phi[\phi(x),t],
\end{eqnarray}
so that  quantum Hamiltonian operator is given by
\begin{eqnarray}\label{23}
H=\int
dx\left[-\frac{\hbar^2}{2}\frac{\delta}{\delta\phi(x)}\frac{\delta}{\delta\phi(x
)}+E[\phi]\right].
\end{eqnarray}
In the field representation the matrix elements of evolution
operator are given by
\begin{eqnarray}\label{24}
\begin{array}{cc}
G(\phi^{(f)}(x),\phi^{(i)}(x),T)=\langle\phi^{(f)}|e^{-\frac{iT}{\hbar}H}|\phi^{
(i)}\rangle\\
\\
=\int D[\phi(x,t)]\exp{(\frac{-i}{\hbar}S[\phi])},
\end{array}
\end{eqnarray}
where the initial conditions are those of static kink solutions of
classical equations where $\phi^{(i)}(x,0)=\phi_{k}(x)$,
$\phi^{(f)}(x,T)=\phi_{k}(x)$. In  semi-classical picture, we are
interested in loop expansion for evolution operator up to the  first
quantum correction
\begin{eqnarray}\label{25}
\begin{array}{cc}
G(\phi^{(f)}(x),\phi^{(i)}(x),\beta)=\\
\\
\exp{(-\frac{\beta}{\hbar}E[\phi_k])}
Det^{-\frac{1}{2}}\left[-\partial^2_{\tau}+P\Delta\right](1+{\cal
O}(\hbar)),
\end{array}
\end{eqnarray}
where we use analytic continuation to Euclidean time,
$t=-i\tau$,$T=-i\beta$, and $\Delta$ is the differential operator
\begin{eqnarray}\label{26}
 \Delta=-\frac{d^2}{d
 x^2}+\frac{d^2V}{d\phi^2}\mid_{\phi=\phi_k},
\end{eqnarray}
P is the projector over the strictly positive part of spectrum of
$\Delta$
\begin{eqnarray}\label{27}
\Delta\xi_n(x)=\omega_n^2\xi_n(x),\,\,\,\,\omega^{2}_{n}\,\,
\epsilon \,\, Spec(\Delta)= Spec(P\Delta)+\{0\}.
\end{eqnarray}
We write  functional determinant in  the form
\begin{eqnarray}\label{28}
Det\left[-\frac{\partial^2}{\partial\tau^2}+\Delta\right]=\prod_{n}det\left[-\frac{\partial^2}{
\partial\tau^2}+\omega^{2}_{n}\right].
\end{eqnarray}
All  determinants in  infinite product correspond to harmonic
oscillators of frequency $\omega_n$. On the other hand, it is well
known that \cite{10}
\begin{eqnarray}\label{29}
\begin{array}{cc}
det\left(-\frac{\partial^{2}}{\partial \tau^2}+\omega^2_n
\right)^{-\frac{1}{2}}
= \prod_{j=1}^N \left(\frac{j^2\pi^2}{\beta^2}+\omega_n^2\right)^{-\frac{1}{2}}\\
\\
=
\prod_j\left(\frac{j^2\pi^2}{\beta^2}\right)^{-\frac{1}{2}}\prod_j\left(1+\frac{\omega^2_n\beta^2}{j^2\pi^2}\right)^{-\frac{1}{2}}.
\end{array}
\end{eqnarray}
The first product dose not depend on $\omega_n$ and combines with
the Jacobian and other factors we have collected into a single
constant. The second factor has the limit
$\left[\frac{\sinh(\omega_n\beta)}{\omega_n\beta}\right]^{-\frac{1}{2}}$,
and thus, with an appropriate normalization, we obtain for large
$\beta$
\begin{eqnarray}\label{30}
\begin{array}{cc}
G(\phi^{(f)}(x),\phi^{(i)}(x),\beta)\cong\\
\\
\exp{(-\frac{\beta}{\hbar}E[\phi_k])}\prod_{n}(\frac{\omega_n}{\pi\hbar})^{\frac{1
}{2}}\exp{\left(-\frac{\beta}{2}\sum_{n}\omega_n(1+{\cal
O}(\hbar))\right)}
\end{array}
\end{eqnarray}
where eigenvalues in the kernel of $\Delta$ have been excluded.
Interesting eigenenergy wave functionals
\begin{eqnarray}\label{31}
H\Phi_j[\phi_k(x)]=\varepsilon_j\Phi_j[\phi_{k}(x)]
\end{eqnarray}
we have an alternative expression for $G_E$ for
$\beta\rightarrow\infty$.
\begin{eqnarray}\label{32}
\begin{array}{cc}
G(\phi^{(f)}(x),\phi^{(i)}(x),\beta)\cong\\
\\
\Phi^{*}_{0}[\phi_k(x)]\Phi_{0}[\phi_k(x)]\exp{(-\beta\frac{\varepsilon_0}{\hbar
})},
\end{array}
\end{eqnarray}
and, therefore, from (\ref{30}) and (\ref{32}) we obtain
\begin{eqnarray}\label{33}
\varepsilon_0^k=E[\phi_k]+\frac{\hbar}{2}\sum_{\omega^{2}_{n}>0}\omega_n+{\cal
O}(\hbar^2),
\end{eqnarray}
\begin{eqnarray}\label{34}
|\Phi_0[\phi_k(x)]|^2=Det^{\frac{1}{4}}\left[\frac{P\Delta}{\pi^2\hbar^2}\right],
\end{eqnarray}
as the Kink ground state energy and wave functional up to One-Loop
order.\\
 If we define
the generalized zeta function
\begin{eqnarray}\label{35}
\zeta_{P\triangle}(s)=Tr(P\Delta)^{-s}=\sum_{\omega^2_n>0}\frac{1}{(\omega^2_n)^
s},
\end{eqnarray}
 associated to  differential operator $P\triangle$, then
\begin{eqnarray}\label{36}
\begin{array}{cc}
\varepsilon_0^k=E[\phi_k]+\frac{\hbar}{2}Tr(P\Delta)^{\frac{1}{2}}+{\cal
O }(\hbar^2)
=\\
\\E[\phi_k]+\frac{\hbar}{2}\zeta_{P\Delta}(-\frac{1}{2})+{\cal
O} (\hbar^2).
\end{array}
\end{eqnarray}
The eigenfunction of $\Delta$ is a basis for  quantum fluctuations
around kink background, therefore sum of the associated zero-point
energies encoded in $\zeta_{P\Delta}(-\frac{1}{2})$ in (\ref{35}) is
infinite. According to zeta function regularization procedure,
energy and mass renormalization prescription, the renormalized kink
energy in semi-classical limit becomes \cite{11}
\begin{eqnarray}\label{37}
\begin{array}{cc}
\varepsilon^k(s)=E[\phi_k]+\Delta M_k +{\cal O}(\hbar^2)
=E[\phi_k]+\\
\\
\lim_{s\rightarrow\frac{-1}{2}}[\delta_1\varepsilon^k(s)+\delta_
2\varepsilon^k(s)]+{\cal O}(\hbar^2),
\end{array}
\end{eqnarray}
where
\begin{eqnarray}\label{38}
\begin{array}{cc}
\delta_1\varepsilon^k(s)= \frac{\hbar}{2}\mu^{2s+1}[\zeta_{P\Delta}(s)-\zeta_{\nu}(s)],\\
\\
\delta_2\varepsilon^k(s)=
\lim_{L\rightarrow\infty}\frac{\hbar}{2L}\mu^{2s+1}\frac{\Gamma(s+1)}{\Gamma(s)
}
\zeta_\nu(s+1)\times\\
\\
\int_{-\frac{L}{2}}^{\frac{L}{2}}
dx\left[\frac{d^2V}{d\phi^2}|_{\phi_k}-\frac{d^2V}{d\phi^2}|_{\phi\nu}\right].
\end{array}
\end{eqnarray}
Here $\phi_\nu$ is a constant minimum of potential $V(\phi)$,
$E[\phi_k]$ is the corresponding classical energy where $\mu$ has
the unit  $length^{-1}$ dimension, introduced to make the terms in
$\delta_1\varepsilon^k(s)$ and $\delta_2\varepsilon^k(s)$
homogeneous from a dimensional point of view and $\zeta_\nu$ denoted
zeta function associated with vacuum $\phi_v$.\\
Now we explain very briefly how one can calculate  zeta function of
an operator though  heat kernel method. We introduce  generalized
Riemann zeta function of operator A by
\begin{eqnarray}\label{39}
\zeta_{A}(s)=\sum_{n}\frac{1}{|\lambda_n|^s},
\end{eqnarray} where
$\lambda_n$ are eigenvalues of operator $A$.
  On the other hand, $\zeta_A(s)$ is
 the Mellin transformation of heat kernel $G(x,y,t)$ which satisfies
 the following heat diffusion equation
 \begin{eqnarray}\label{40}
 A G(x,y,t)=-\frac{\partial}{\partial t}G(x,y,t),
 \end{eqnarray}
 with an initial condition $G(x,y,0)=\delta(x-y)$. Note that
 $G(x,y,t)$ can be written in terms of its spectrum
 \begin{eqnarray}\label{41}
 G(x,y,t)=\sum_{n}e^{-\lambda_n t}\psi_n^{*}(x)\psi_n(y),
 \end{eqnarray}
 and as usual, if the spectrum is continues, one should integrate it. From relation (\ref{39}), (\ref{40}) and (\ref{41}) it is clear that
\begin{eqnarray}\label{42}
 \zeta_A(s)=\frac{1}{\Gamma(s)}\int_{0}^{\infty}d\tau\tau^{s-1}\int_{-\infty}^{
 \infty}G(x,x,\tau)dx.
 \end{eqnarray}
 Hence, if we know the associated Green function of an operator,
 we can calculate the generalized zeta function corresponding
 to that operator. In the next section, we use
the shape invariance property of the potential and the heat kernel
method to obtain the quantum corrections to the kink masses.

\section{Renormalized ground state energy of the
Kink solution in de Sitter universe}

Comparison of the two previous sections reveals that
the system of closed de Sitter universe may be equivalent
to the classical configuration space of the static field
$\Phi(x)$. Therefore, the one-loop quantum corrections to the mass quantity of closed de Sitter universe, namely $m=3\pi c^2/(2G)$, is equivalent to the one-loop quantum corrections to the kink mass or ground state energy of the scalar field $\Phi(x)$. We call this procedure as the one-loop quantum corrections to the ground state energy of the Kink-like solution of closed  de Sitter universe. In other words, what is physically meant by the one-loop quantum corrections to the ground state energy in  de Sitter universe is nothing but the
one-loop quantum corrections to the mass quantity $m=3\pi c^2/(2G)$ in  de Sitter universe which resembles the kink's mass or energy in the equivalent system of the classical static field $\Phi(x)$ configuration.  

Using the techniques of the previous section implemented on the scalar field
theory, we now compute the one-loop quantum correction to the ground state energy of the Kink-like solution of de Sitter universe. To
this end, we need the spectrum of differential operator
(\ref{26}) and the corresponding vacuum. According to the previous
section, the operator (\ref{26}) which acts on the eigenfunctions becomes
\begin{eqnarray}\label{44}
\Delta_{l+h} = \frac{mc^2}{\hbar}\left(-\frac{d^2}{dx^2} +l^2-\frac{l(l+1)}{\cosh^2 x}+h\right),
\end{eqnarray}
where $x=\sqrt{m}c\tau$, $h=1-2l$ and  $l=2$. The combination $\frac{mc^2}{\hbar}$
is a necessary result of the well known
fact that in every quantum mechanical problem of gravity where $\hbar$ appears,
$m$ is also expected to appear \cite{Sakurai}.
  
Also the operator acting on the vacuum has the following form
\begin{eqnarray}\label{45}
\Delta_{l+h}(0)=\frac{mc^2}{\hbar}\left(-\frac{d^2}{dx^2}+l^2+h\right).
\end{eqnarray}
Note that we have the constant shift $h$ in the spectrum that we add it in
our calculations latter (see Eq.({\ref{71})), also since we have  $\zeta_{\sigma\Delta}(s)=|\sigma|^{-s}\zeta_\Delta(s)$
so we ignore $\sigma=\frac{mc^2}{\hbar}$ here in  our calculations and we add it at last steps.
In the reminding of this section, to obtain the spectrum of
(\ref{44}) we will use the shape invariance property. First we
review briefly concepts
that we will use.\\
Consider the following one-dimensional bound-state Hamiltonian
\begin{eqnarray}\label{46}
H= -\frac{d^2}{dx^2}+U(x), \hspace{2cm} x\in I \subset {\cal R}
\end{eqnarray}
where $I$ is the domain of  $x$ and $U(x)$ is a real function of
$x$, which can be singular only in the boundary points of the
domain. Let us denote by $E_n$ and $\psi_n(x)$ the eigenvalues and
eigenfunctions of $H$ respectively. We use factorization method
which consists of writing Hamiltonian as the product of two first
order mutually adjoint differential operators $A$ and $A^\dagger$.
If the ground state eigenvalue and eigenfunctions are known, then
one can factorize Hamiltonian (\ref{46}) as
\begin{eqnarray}\label{47}
H=A^\dagger A +E_0,
\end{eqnarray}
where $E_0$ denotes the ground-state eigenvalue,
\begin{eqnarray}\label{48}
\begin{array}{lll}
A=\frac{d}{dx}+W(x),\\
\\
A^\dagger=-\frac{d}{dx}+W(x),
\end{array}\
\end{eqnarray}
and
\begin{eqnarray}\label{49}
W(x)=-\frac{d}{dx}\ln(\psi_0).
\end{eqnarray}
Supersymmetric quantum mechanics (SUSY QM) begins with a set of two
matrix operators, known as supercharges
\begin{equation}
Q^+ =\left(\begin{array}{cc} 0 & A^\dagger\\
\\0 & 0 \end{array}\right)_, \hspace{2cm} Q^- =\left(\begin{array}{cc} 0 & 0\\
\\A & 0 \end{array}\right)_.\label{50}
\end{equation}
This operators form the following superalgebra \cite{14}
\begin{eqnarray}\label{51}
\{Q^+,Q^-\}=H_{SS}, \hspace{1cm} [H_{SS},Q^{\pm}]=(Q^\pm)^2=0,
\end{eqnarray}
where SUSY Hamiltonian $H_{SS}$ is defined as
\begin{equation}
H_{SS}=\left(\begin{array}{cc} A^\dagger A & 0\\
\\0 & AA^\dagger \end{array}\right)=\left(\begin{array}{cc} H_1 & 0\\
\\0 & H_2 \end{array}\right)_.\label{52}
\end{equation}
In terms of the Hamiltonian supercharges
\begin{eqnarray}\label{53}
\begin{array}{ccc}
Q_1=\frac{1}{\sqrt{2}}(Q^++Q^-),\\
\\
Q_2=\frac{1}{\sqrt{2i}}(Q^+-Q^-),
\end{array}
\end{eqnarray}
the superalgebra takes the form
\begin{eqnarray}\label{54}
\{Q_i,Q_j\}=H_{SS}\delta_{ij}, \hspace{0.5cm}[H_{SS},Q_i]=0,
\hspace{0.5cm} i,j=1,2.
\end{eqnarray}
The operators $H_1$ and $H_2$
\begin{eqnarray}\label{55}
\begin{array}{cc}
H_1= A^\dagger  A = -\frac{d^2}{dx^2} +U_1=-\frac{d^2}{dx^2}+W^2-\frac{dW}{dx},\\
\\
H_2=AA^\dagger = -\frac{d^2}{dx^2} + U_2
=-\frac{d^2}{dx^2}+W^2+\frac{dW}{dx},
\end{array}
\end{eqnarray}
are called SUSY partner Hamiltonians and the function $W$ is called
the superpotential. Now, let us denote by $\psi^{(1)}_{\,\,\, l}$
and $\psi^{(2)}_{\,\,\, l}$ the eigenfunctions of $H_1$ and $H_2$
with eigenvalues $E^{(1)}_{l}$ and $E^{(2)}_l$, respectively. It is
easy to see that the eigenvalues of the above Hamiltonians are
positive and isospectral, i.e., they have almost the same energy
eigenvalues, except for the ground state energy of $H_1$. According
to the \cite{14}, their energy spectra are related as
\begin{eqnarray}\label{56}
\begin{array}{cccc}
E_l=E^{(1)}_l+E_0, & E^{(1)}_0=0, & \psi_l=\psi^{(1)}_l,& l=0,1,2,.., \\
\\
E^{(2)}_l=E^{(1)}_{l+1},\\
\\
\psi^{(2)}_l = [E^{(1)}_{l+1}]^{-\frac{1}{2}}A\psi^{(1)}_{l+1},\\
\\
\psi^{(1)}_{l+1} =
[E^{(2)}_{l}]^{-\frac{1}{2}}A^\dagger\psi^{(2)}_{l}.
\end{array}
\end{eqnarray}
Therefor if the eigenvalues and eigenfunctions of $H_1$ were known,
one could immediately derive the spectrum of $H_2$. However the
above relations only give the relationship between the eigenvalues
and eigenfunctions of the two partner Hamiltonians. A condition of
an exactly solvability is known as the shape invariance condition.
This condition means the pair of SUSY partner potentials
$U_{1,2}(x)$ are similar in shape and differ only in the parameters
that appears in them \cite{15}
\begin{eqnarray}\label{57}
U_2(x;a_1)=U_2(x;a_2)+{\cal R}(a_1),
\end{eqnarray}
where $a_1$ is a set of parameters and $a_2$ is a function of $a_1$.
Then the eigenvalues of $H_1$ are given by
\begin{eqnarray}\label{58}
E^{(1)}_l={\cal R}(a_1)+{\cal R}(a_2)+...+{\cal R}(a_l),
\end{eqnarray}
and the corresponding eigenfunctions are
\begin{eqnarray}\label{59}
\psi_l=\prod^{l}_{m=1}\frac{A^\dagger(x;a_m)}{\sqrt{E_m}}\psi_0(x;a_{l+1}).
\end{eqnarray}
The shape invariance condition (\ref{57}) can be rewritten in terms
of the factorization operators defined in equation (\ref{48})
\begin{eqnarray}\label{60}
A(x;a_1)A^\dagger(x;a_1) = A^\dagger(x;a_2) A(x;a_2)+{\cal R}(a_1),
\end{eqnarray}
where $a_2=f(a_1)$. Now we are ready to obtain spectra of $\Delta_l$
operator defined in (\ref{28}). For a given eigenspectrum of $E_l$,
we introduce the following factorization operators
\begin{eqnarray}\label{61}
\begin{array}{cc}
A_l=\frac{d}{dx}+l\tanh(x),\\
\\
A^\dagger_l=-\frac{d}{dx}+l\tanh(x),
\end{array}
\end{eqnarray}
the operator $\Delta_l$ can be factorized as
\begin{eqnarray}\label{62}
\begin{array}{cc}
 A^\dagger_l(x)A_l(x)\psi^{(1)}_n(x)=E^{(1)}_n\psi^{(1)}_n(x),\\
 \\
 A_l(x)A^\dagger_l(x)\psi^{(2)}_n(x)=E^{(2)}_n\psi^{(2)}_n(x).
 \end{array}
 \end{eqnarray}
 Therefor for a given $l$, its first bounded excited state can be obtained
 from the ground state of $l-1$ and consequently the excited state $m$ of
 a given $l$, $\psi_{l,m}(x)$, using (\ref{59}) can be written as
 \begin{eqnarray}\label{63}
 \psi_{l,m}(x) = \sqrt{\frac{2(2m-1)!}{\Pi_{j=1}^mj(2l-j)}}\frac{1}{2^m(m-1)!}A^\dagger_l(x)
 A^\dagger_{l-1}(x)...A^\dagger_{m+1}(x)\frac{1}{\cosh^m(x)},
 \end{eqnarray}
 with eigenvalue $E_{l,m}=m(2l-m)$. Obviously its ground state with $E_{l,0}=0$
 is given by $\psi_{l,0}\propto \cosh^{-l}(x)$. Also its continuous spectrum
 consists of
 \begin{eqnarray}\label{64}
 \psi_{l,k}(x)=\frac{A^\dagger_{l}(x)}{\sqrt{k^2+l^2}}\frac{A^\dagger_{l-1}(x)}{\sqrt{k^2+(l-1)^2}}
 ...\frac{A^\dagger_{1}(x)}{\sqrt{k^2+1}}\frac{e^{ikx}}{\sqrt{2\pi}},
 \end{eqnarray}
 with eigenvalues $E_{l,k}=l^2+k^2$ with following  normalization condition  \begin{eqnarray}\label{65}
 \int_{-\infty}^\infty \psi^*_{l,k}(x)\psi_{l,k'}(x)dx=\delta(k-k').
 \end{eqnarray}
 Therefor, using equations (\ref{40}), (\ref{41}), (\ref{63}) and (\ref{64})
 we find
 \begin{eqnarray}\label{66}
 G_{\Delta_l(0)}(x,y,\tau)=\frac{e^{-l^2\tau}}{2\sqrt{\pi \tau}}e^{-(x-y)^2/4\tau},
 \end{eqnarray}
 and
 \begin{eqnarray}\label{67}
 \begin{array}{cc}
 G_{\Delta_l}(x,y,\tau)=\sum_{m=1}^{l-1}\psi^*_{l,m}(x)\psi_{l,m}(y)e^{-m(2l-m)\tau}\\
 \\
 +\int_{-\infty}^\infty  \frac{dk}{2\pi}\frac{e^{-(l^2+k^2)\tau}}{\prod_{m=1}^l(k^2+m^2)}\left(\prod_{m=1}^lA^\dagger_m(x)e^{ikx}\right)^*
\left(\prod_{m=1}^lA^\dagger_m(y)e^{iky}\right).
\end{array}
\end{eqnarray}

In the case of de Sitter spacetime we left with $l=2$ and then using
(\ref{42}) we have
\begin{eqnarray}\label{69}
\begin{array}{cc}
\zeta_{\delta_2}(s)-\zeta_{\delta_2(0)}(s)=3^{-s}-\frac{3}{\pi}\int_{-\infty}^\infty
\frac{dk}{(k^2+4)^{s+1}}=\\
\\
3^{-s}-\frac{3}{\sqrt{\pi}}2^{-(2s+1)}\frac{\Gamma(s+\frac{1}{2})}{\Gamma(s+1)}.
\end{array}
\end{eqnarray}

Consequently we have
\begin{eqnarray}\label{71}
\begin{array}{cc}
\delta_1\varepsilon^k(s)=\frac{\hbar}{2}\mu^{2s+1}(\frac{mc^2}{\hbar})^{-2s}\left[\zeta_{P\Delta_{l+h}}(s)-\zeta_{\Delta_{(v)l+h}}(s)\right]_{l=2}=\\
\\
\frac{\hbar}{2}\mu^{2s+1}(\frac{mc^2}{\hbar})^{-2s}
\left(3^{-s}-\frac{3}{\sqrt{\pi}}2^{-(2s+1)}\frac{\Gamma(s+\frac{1}{2})}{\Gamma(s+1)}]\right).
\end{array}
\end{eqnarray}
Also we obtain
\begin{eqnarray}\label{72}
\begin{array}{cc}
\delta_2\varepsilon^k(s)= \lim_{L\rightarrow
\infty}\frac{\hbar}{2L}(\frac{mc^2}{\hbar})^{-2s}\mu^{2s+1}\frac{\Gamma(s+1)}{\Gamma(s)}
\zeta_{\Delta_l+h(v)}(s+1)|_{l=2}\int_{-\frac{L}{2}}^\frac{L}{2}dx(-6\cosh^{-2}(x))\\
\\
=
-\frac{3\hbar}{\sqrt{\pi}}\mu^{2s+1}(\frac{mc^2}{\hbar})^{-2s}\frac{\Gamma(s+\frac{1}{2})}{\Gamma(s)}.
\end{array}
\end{eqnarray}
Finally we have
\begin{eqnarray}\label{73}
\begin{array}{ccc}
\lim_{s\rightarrow
-\frac{1}{2}}(\delta_1\varepsilon^k(s)+\delta_2\varepsilon^k(s))=\frac{\hbar}{2}\mu^{2s+1}(\frac{mc^2}{\hbar})^{-2s}\left(3^{-s}-\frac{3}{\sqrt{\pi}}\frac{\Gamma(s+\frac{1}{2})}{\Gamma(s+1)}-\frac{6}{\sqrt{\pi}}\frac{\Gamma(s+\frac{1}{2})}{\Gamma(s)}\right)\\
\\
=\frac{\hbar}{2}\frac{mc^2}{\hbar}\left(\sqrt{3}-\frac{3}{\pi}\right).
\end{array}
\end{eqnarray}
At last, we find the one-loop correction to the
kink energy (mass) as
\begin{eqnarray}\label{74}
\varepsilon^k=mc^{2}+\frac{mc^2}{2}(\sqrt{3}-\frac{3}{\sqrt{\pi}}),
\end{eqnarray}
Now, the renormalized gravitational constant $G_{one-loop}$ is obtained through
$G=\frac{\frac{3}{2}\pi c^4}{mc^2}$ as
\begin{eqnarray}\label{75}
G_{one-loop}=\frac{G}{1+\frac{1}{2}(\sqrt{3}-\frac{3}{\sqrt{\pi}})},
\end{eqnarray}
which indicates that the one-loop renormalized gravitational constant is smaller than the original gravitational constant.

\section{Conclusion}

The shape invariance property of the fluctuation operator is of
particular importance in order to find the exact one-loop quantum
correction to the mass of kink solutions, namely the instanton
solutions of classical field equations. The (1+1)-dimensional
Sine-Gordon and $\phi^4$ field theories are some of these examples.
This kind of potential occurs in different fields of physics
particularly in quantum gravity and cosmology. The de Sitter
spacetime is one of those examples whose quantum cosmology reveals
the shape invariance property in the action. 

In this paper, we have shown that the system of closed de Sitter universe is equivalent to the classical configuration space of the static field
$\Phi(x)$. Therefore, we expect kink-like solutions in the system of closed de Sitter universe. The one-loop quantum corrections to the mass-like quantity of closed de Sitter universe, namely $m=3\pi c^2/(2G)$, is therefore equivalent to the computation of the one-loop quantum corrections to the kink mass or ground state energy of the scalar field $\Phi(x)$. In fact,
what is physically meant by the one-loop quantum corrections to the ground state energy in de Sitter universe, in the present paper, is nothing but the one-loop quantum corrections to the mass quantity $m=3\pi c^2/(2G)$ in de Sitter universe which plays the role of kink's mass in the equivalent system of the classical static field $\Phi(x)$ configuration. We have therefore computed the quantum corrections to the ground state energy of the Kink-like solution in the closed de Sitter universe. From mass-energy equivalence principle we know that any corrections to the energy is equivalent to the corresponding corrections to an equivalent mass quantity. In de Sitter cosmology this mass quantity coincides with the gravitational constant as $m=3\pi c^2/(2G)$. Therefore, it is shown that the one-loop quantum corrections to the ground state energy of the Kink-like solution in the closed de Sitter universe may
be interpreted as the renormalization of the gravitational constant which turns out to be smaller than the original gravitational constant. The obtained correction may become more viable and important whenever we study the other interesting aspects of quantum cosmology of closed de Sitter universe. For example, this correction
may have considerable impact in evaluation of the tunneling rate from "nothing" to a closed de Sitter universe. In fact, since the tunneling rate depends on the classical action which itself is dependent on the gravitational constant, then any renormalization on the gravitational constant will alter the tunneling rate. 

We feel it necessary to compare and discuss on the present semi classical
one-loop quantum {\it cosmological} corrections
and the well known one-loop quantum {\it gravitational} corrections. In the latter approach, for instance in de Sitter spacetime, one usually uses of zeta-function regularization technique and heat-kernel methods to compute the leading part of the one-loop contribution to the effective action. As a result, addition of this contribution to the classical action leads to the one-loop effective action in the large-distance limit as \cite{15.15}
\begin{eqnarray}\nonumber
\Gamma_{eff}&=& S+\Gamma(1)
\\
\nonumber
&=&\int d^4x\sqrt{g}[\Lambda(8\pi G)^{-1}+\beta_{\Lambda}|\Lambda|^2\log(|\Lambda|\mu^{-2})]
\\
\nonumber
&-&\int d^4x\sqrt{g}R[(16\pi G)^{-1}+\beta_{G}|\Lambda|\log(|\Lambda|\mu^{-2})],
\end{eqnarray}
where the effective or induced Newton and cosmological constants are given by
$$
\Lambda_{eff}=\Lambda\frac{1+\kappa\beta_{\Lambda} 8 \pi G|\Lambda| \log(|\Lambda|\mu^{-2})}{1+ \beta_{G}16\pi G|\Lambda| \beta_{\Lambda}\log(|\Lambda|\mu^{-2})},
$$
$$
(G\Lambda)_{eff}=(G\Lambda)\frac{1+\kappa\beta_{\Lambda} 8 \pi G|\Lambda| \log(|\Lambda|\mu^{-2})}{[1+ \beta_{G}16\pi G|\Lambda| \beta_{\Lambda}\log(|\Lambda|\mu^{-2})]^2}.
$$
In fact, for any background like de Sitter, one should calculate vacuum energy (effective) action which yields effective gravitational and cosmological constants. It is then impossible to calculate only gravitational constant
unless we prove that the induced cosmological constant is zero by some (yet
unknown) mechanisms.

In the present paper, however, rather than the effective action
approach we are just dealing with the semi classical correction on the kink-like solution in de Sitter universe and no such corrections are imposed by effective action on the gravitational and cosmological constants. Thus, we just obtain a semi classical correction of mass which may be {\it interpreted} as a correction to the gravitational constant due to the relation $m=3\pi c^2/(2G)$.
The cosmological constant here appears just as the inverse squared of the
minimum radius of the universe after tunneling from nothing, namely $\Lambda=\frac{3}{a_0^2}$, which indicates the bouncing point or the width of the potential barrier in the Wheeler-DeWitt equation, while the gravitational constant plays the role of the mass of kink-like solution. Therefore, it is reasonable to expect that the mass term, namely gravitational constant, as a dynamical quantity
bears quantum correction but the cosmological constant as a parameter in the potential bears no such a correction.

\section*{Acknowledgment}
The authors would like to thank the anonymous referee for the useful comments
and Prof. S. D. Odintsov for the valuable hints and discussion.
This work  has been supported by the Research office of Azarbaijan
University of Tarbiat Moallem, Tabriz, Iran.

\end{document}